\documentclass[sn-mathphys,Numbered]{sn-jnl}

\usepackage{graphicx}%
\usepackage{multirow}%
\usepackage{amsmath,amssymb,amsfonts}%
\usepackage{amsthm}%
\usepackage{mathrsfs}%
\usepackage[title]{appendix}%
\usepackage{xcolor}%
\usepackage{textcomp}%
\usepackage{manyfoot}%
\usepackage{booktabs}%
\usepackage{algorithm}
\usepackage{algpseudocode}
\usepackage{algorithmicx}
\usepackage{algpseudocode}%
\usepackage{listings}%
\usepackage{subfigure}
\usepackage{longtable}
\usepackage{helvet}  
\usepackage{courier}  
\usepackage{url}
\usepackage{caption} 
\usepackage{comment}
\usepackage{color}
\usepackage{booktabs}
\usepackage{color}
\usepackage{bm}
\usepackage{siunitx} 

\usepackage{times}  

\usepackage{enumitem}
\usepackage{caption}
\usepackage{lipsum}
\usepackage{threeparttable}
\usepackage{mathtools} 

\usepackage{tikz}
\usetikzlibrary{arrows.meta}

\usepackage{verbatim}

\usepackage{newfloat}

\theoremstyle{thmstyleone}%
\theoremstyle{thmstyletwo}%

\theoremstyle{thmstylethree}%

\begin{document}

\title[Article Title]{Integrating Deep Learning and Synthetic Biology: A Co-Design Approach for Enhancing Gene Expression via N-terminal Coding Sequences}

\author*[1]{\fnm{Zhanglu} \sur{Yan}}\email{zhangluyan@comp.nus.edu.sg}
\equalcont{These authors contributed equally to this work.}

\author[2]{\fnm{Weiran} \sur{Chu}}\email{7220201036@stu.jiangnan.edu.cn}
\equalcont{These authors contributed equally to this work.}

\author[2]{\fnm{Yuhua} \sur{Sheng}}\email{6220210017@stu.jiangnan.edu.cn}
\author[1]{\fnm{Kaiwen} \sur{Tang}}\email{tang\_kaiwen@u.nus.edu}
\author[3]{\fnm{Shida} \sur{Wang}}\email{e0622338@u.nus.edu}

\author*[2]{\fnm{Yanfeng} \sur{Liu}}\email{yanfengliu@jiangnan.edu.cn}

\author[1]{\fnm{Weng-Fai} \sur{Wong}}\email{wongwf@comp.nus.edu.sg}

\affil[1]{\orgdiv{School
of Computing}, \orgname{National University of Singapore}, \orgaddress{\street{21 Lower Kent Ridge Road}, \city{Singapore}, \postcode{119077}, \country{Singapore}}}

\affil[2]{\orgdiv{Science Center for Future Foods}, \orgname{Jiangnan University}, \orgaddress{\street{No. 1800, Lihu Avenue}, \city{Wuxi}, \postcode{214122}, \country{China}}}

\affil[3]{\orgdiv{Department of Mathematics}, \orgname{National University of Singapore}, \orgaddress{\street{21 Lower Kent Ridge Road}, \city{Singapore}, \postcode{119077}, \country{Singapore}}}


\abstract{




N-terminal coding sequence (NCS) influences gene expression by impacting the translation initiation rate. The NCS optimization problem is to find an NCS that maximizes gene expression. The problem is important in genetic engineering. However, current methods for NCS optimization such as rational design and statistics-guided approaches are labor-intensive yield only relatively small improvements. This paper introduces a deep learning/synthetic biology co-designed few-shot training workflow for NCS optimization. Our method utilizes $k$-nearest encoding followed by word2vec to encode the NCS, then performs feature extraction using attention mechanisms, before constructing a time-series network for predicting gene expression intensity, and finally a direct search algorithm identifies the optimal NCS with limited training data. We took green fluorescent protein (GFP) expressed by {\em Bacillus subtilis} as a reporting protein of NCSs, and employed the fluorescence enhancement factor as the metric of NCS optimization. Within just six iterative experiments, our model generated an NCS (MLD$_{62}$) that increased average GFP expression by 5.41-fold, outperforming the state-of-the-art NCS designs.
Extending our findings beyond GFP, we showed that our engineered NCS (MLD$_{62}$) can effectively boost the production of N-acetylneuraminic acid by enhancing the expression of the crucial rate-limiting {\em GNA1} gene, demonstrating its practical utility. We have open-sourced our NCS expression database and experimental procedures for public use.}




\maketitle

\section{Introduction}\label{sec1}

\begin{figure*}[ht]
    \centering
    \includegraphics[width = 0.85\linewidth]{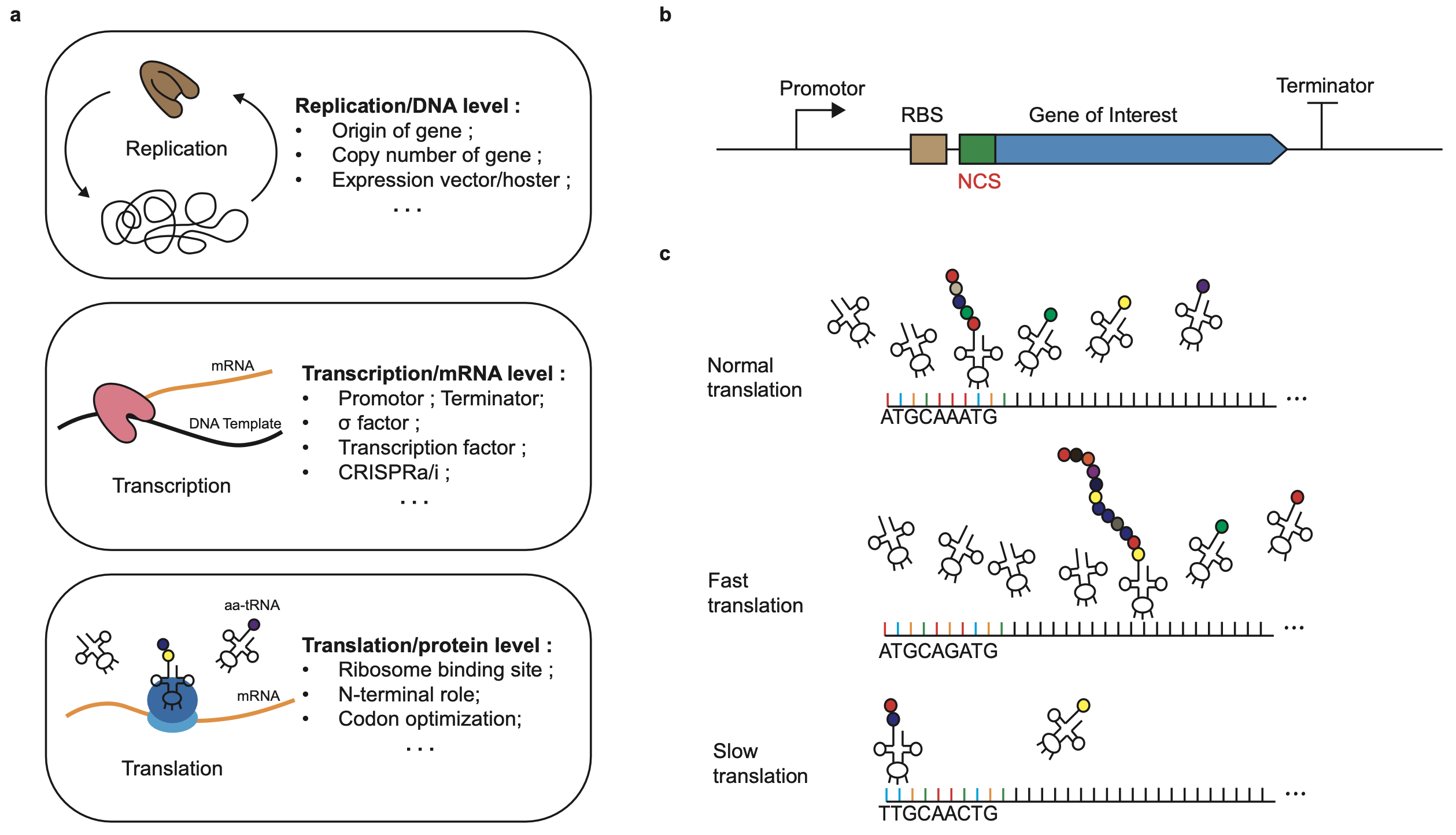}
        \caption{Gene adjusting methods.}
    \label{fig:workflow_1_}
\begin{tablenotes}

\item {(a) Gene expression regulation toolkit; (b) gene expression cassette; (c) NCS translation rate.}
\end{tablenotes}
\end{figure*}

Precise control of gene expression is essential in synthetic biology~\cite{horton2023short, gil2020regulation, bosch2021genome}. Existing strategies, as shown in Figure~\ref{fig:workflow_1_}(a), have focused on modulating gene expression at different stages~\cite{fu2022operator, lu2019crispr, ding2020programmable,lv2023crispr}. However, each level of manipulation presents its own set of challenges. For example, modifying replication levels always leads to lower gene expression increases than at other levels, while adjusting transcription will lead to outcomes with low robustness and high variance~\cite{yang2017characterization}. In contrast, adjustment at the translation level, which is used in this paper, can ensure increased, stable gene expression~\cite{tian2019synthetic,fredrick2010sequence}.

Numerous genetic regulatory strategies, including ribosome binding site (RBS) screening, codon optimization, and N-terminal role adjustment, are commonly used for precise control of gene expression intensity at the translation level~\cite{tian2020titrating,zhao2021directed,stork2021designing}. 
Among these, {\em N-terminal coding sequences} (NCS) influence gene expression by impacting the binding and extension efficiency between ribosome and mRNA during the translation initiation stage~\cite{cambray2018evaluation, goodman2013causes, kudla2009coding, espah2017precise}. This demonstrates the potential of NCS for refined regulation of gene expression. However, the impact of NCS on gene expression has remained largely theoretical, with current computational tools unable to predict the precise expression intensity. This limitation impedes the utilization of NCS as a regulatory element for modulating metabolic pathway expressions.

\begin{figure*}[ht]
    \centering
    \includegraphics[width = 0.95\linewidth]{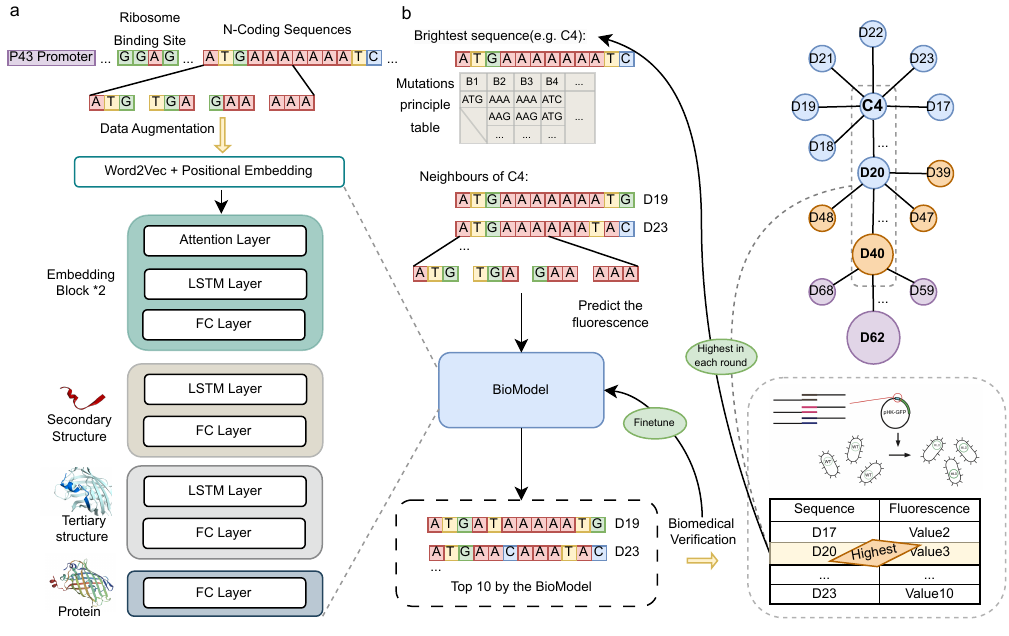}
        \caption{Workflow of our training methods.}
    \label{fig:workflow_2}
\begin{tablenotes}

\item {(a) NCS encoding methods and model design;
(b) Deep learning/ synthetic biology co-designed few-shot training workflow}
\end{tablenotes}
\end{figure*}

Traditional approaches for studying expression elements, such as manual random mutation ~\cite{wang2022model} and rational artificial gene design ~\cite{tian2019synthetic}, tend to be labor-intensive or inefficient. Consequently, rational design and development of toolkits for modulating gene expression using NCS remains challenging. In this paper, in contrast to traditional designs, we introduce a deep learning/ synthetic biology co-designed few-shot training workflow for gene expression enhancement that is shown in Figure~\ref{fig:workflow_2}.  We start with a $k$-nearest encoding (with $k=3$) to encapsulate information about three adjacent nucleotides, including their values and positional data. Each 3-length nucleotide combination is then mapped into a distinct vector space via the word2Vec (CBOW) algorithm, simultaneously endowing each vector with contextual data during the CBOW model training process~\cite{rong2014word2vec}. 
Furthermore, we introduced an embedding block comprising an attention layer ~\cite{vaswani2017attention} to capture the contextual difference of gene sequences, a {\em long short-term memory} (LSTM) layer~\cite{graves2012long} for addressing long-term dependencies, and a {\em fully connected} (FC) layer to integrate the data.
We employed two such blocks for NCS embedding: the first was tailored to gather nucleotide-specific information, and the second aimed at a broader assimilation of comprehensive NCS data. Finally, we add another FC layer to output the final fluorescence intensity results.

In our experiments, GFP expressed by {\em B. subtilis}. (a commonly used food-grade industrial production strain)~\cite{liu2020chassis,gu2018advances} is used as a reporting protein, and the first 45 base pairs of the GFP gene were selected as the standard length for the NCS. To form an NCS-to-expression intensity dataset, we initially selected 73 genes with varying expression intensity based on the transcriptomic and proteomic data of {\em B. subtilis}. Then, we introduced oligonucleotides to clone the NCS of these 73 genes upstream of the start codon of GFP. We characterized the expression intensity of these 73 NCS variants through fluorescence quantification. The sequences of these NCSs, along with their corresponding fluorescence, constituted the initial training set for our study. Recognizing the challenges of limited and unevenly distributed data, particularly noticeable at higher fluorescence intensities, along with the substantial costs linked to repeated experimentation, we undertook a process of balancing and augmenting the dataset of 73 NCS sequences. Also, we proposed a loss function that gives more weight to the high-level expression NCS training data to bolster the model's efficacy in identifying and forecasting high-expression genetypes.

After training with our loss function on the augmented initial dataset, we use direct search algorithms for NCS genetype with high-level expression, shown in Figure~\ref{fig:workflow_2}(b). Guided by the principles of genetic engineering (at specific locations, a codon can only mutate into certain specific codons, for example, position B1 in mutations principle table of Figure~\ref{fig:workflow_2}(b) can only be ATG), we performed mutations on each codon group of the highest-expressing genetype in our training dataset (for example, C4 genetype in initial training dataset). Each codon had a 20\% chance of being altered into a new, contextually suitable codon. Following this, we identified the top 10 new genetypes with the highest expression intensities as determined by our model, and forwarded them for biological validation. Then, those 10 new genotypes, including their authentic expression intensities obtained by biological validation, were added to the training dataset for following iterations. 


We engineered an NCS (MLD$_{62}$) capable of enhancing GFP expression by an average of 5.41-fold through six rounds of iterative machine learning-driven phenotype validation. {\color{black}This performance exceeds the 2.58-fold increase achieved by the best endogenous NCS ($C_4$) included in our initial dataset. 
It also outperforms the state-of-the-art results reported by Xu~\textit{et al.}~\cite{xu2021rational}, Wang \textit{et al.}~\cite{wang2022model} and Tian \textit{et al.}~\cite{tian2019synthetic}, which we replicated using our experimental conditions for a fairer comparison.}
Our novel approach bypasses the thousands of biological experiments~\cite{xu2021rational,wang2022model}
required by these conventional, labor-intensive techniques, achieving a significantly higher NCS-mediated gene expression regulation in just six experimental cycles, involving only 59 phenotype biology validations.

The main contributions of this paper are:

\begin{itemize}
    
    \item We introduce a novel few-shot training strategy for designing NCS that suits scenarios with limited data. Within six rounds of machine learning-driven phenotype validation, we developed an NCS (MLD$_{62}$) that led to {\color{black}state-of-the-art} increase in GFP expression.

    \item We employ the MLD$_{62}$ to regulate the key rate-limiting gene {\em GNA1} in N-acetylneuraminic acid synthesis. This resulted in a 1.25-fold average enhancement in the performance of this high-value production.
    
    \item We have constructed and made our code and dataset for the above publicly available. We hope this open-source resource will spur ongoing innovation in the field.

\end{itemize}

\section{Results}\label{sec2}




\subsection{NCS encoding} 
In our study, we implemented $k$-nearest encoding to break the NCS into segments of size $k$ for genetic sequence analysis, as outlined in Step 1 of Algorithm~\ref{alg1}. We chose $k=3$ to create segments corresponding to codons, in accordance with the biological principle that each codon consists of a trinucleotide structure. For instance, the $C_4$ genetype sequence ``ATGAAAA..." is segmented into the 3-sequence ``ATG TGA GAA AAA AAA...".  Subsequently, each segment is regarded as an individual ``word" and processed using a {\em continuous bag of words} (CBOW) model via Word2Vec (Step 2 of Algorithm~\ref{alg1}). This method generates vector embeddings, ensuring that segments sharing contextual similarities are grouped nearby in the corresponding vector space. Furthermore, to address the limitations of attention schemes in capturing positional data, we enhance these Word2Vec-derived vectors with positional encodings via a sinusoidal algorithm (Step 3 of Algorithm~\ref{alg1}).



\begin{algorithm}
\caption{NCS encoding }
\label{alg1}
\begin{algorithmic}[1]
\Require NCS$\bm{x}$ with length of $l$; A Word2Vec model $M$; Dimension of $D$ after adapting Word2Vec model; K-nears encoding level $K$
\Ensure Encoded NCS after k-nears encoding $\bm{x}^{kn}$; Encoded NCS after Word2Vec $\bm{x^{w2v}}$; Encoded NCS after position encoding $\bm{x^{p}}$

\begin{minipage}{0.5\textwidth}
\State  {\bf STEP 1 - K-nearest encoding}:
\For{ $k= 0$ to $l-K+1$}
\State $\bm{x^{kn}}_k = \bm{x}[k:k+K]$
\EndFor
\State
\State {\bf STEP 2 - Word2Vec value encoding}
\State $M = M(\bm{x}^{kn})$ // Train Word2Vec
\For{ $k= 0$ to $l-K+1$}
\State $\bm{x^{w2v}}_k = M(\bm{x^{kn}}_k)$
\EndFor
\end{minipage}%
\begin{minipage}{0.5\textwidth}
\State {\bf STEP 3 - Positional encoding}

\For{ $k= 0$ to $l-K+1$}
\For{ $d= 0$ to $D/2-1$}
\State $\bm{x^{p}}_{k,2*d} = \sin\left(\frac{k}{10000^{(2d/D)}}\right)$
\State $\bm{x^{p}}_{k,2*d+1} = \cos\left(\frac{k}{10000^{(2d/D)}}\right)$
\EndFor
\EndFor
\end{minipage}
\State
\State return $\bm{x^{w2v}} +\bm{x^{p}} $

\end{algorithmic}
\end{algorithm}

\subsection{The NCS prediction model} 
We use a neural architecture to predict gene expression intensity that comprise of three distinct blocks. The first block focuses on extracting features from encoded NCS. The second block simulates protein structures such as helices, folds, and loops. The final component is an output layer that predicts the gene expression level.

\begin{itemize}
\item Embedding (contextual feature extraction): This block is designed for precise feature extraction, starting with an attention layer that is adept at identifying contextual relationships among codons. This layer assigns appropriate weight to different inputs, focusing on the most relevant elements. Following this, a long short-term memory (LSTM) layer is incorporated. Its strength lies in processing sequential data and recognizing longer-term dependencies. The final stage of embedding is a fully connected (FC) layer, which integrates the extracted features. Two embedding blocks are utilized here: one targeting detailed trinucleotide-level details and the other focusing on broader NCS sequence characteristics.

\item Processing (protein structure modeling): In this block, we present a series of custom-designed layers, including an LSTM and an FC layer, specifically for modeling protein structures like helices, folds, and loops. The LSTM part manages the temporal aspects of sequences, while the FC layer compiles and interprets the insights gathered from the previous layer, effectively capturing the positional relationships within the protein's structure. Our model has two processing blocks to reflect the secondary and tertiary structural intricacies of proteins.

\item Output: The architectural model ends with one FC layer which is responsible for delivering the final NCS expression intensity.

\end{itemize}

In addition, to train the model, we designed a specialized loss function, which we call \textit{piecewise MSE loss}, to give priority to the learning of high-fluorescence genetypes (those with fluorescence above a threshold, $\upsilon$) by applying a scaling factor of $\beta$. The loss function is as follows:

\begin{equation}
\label{eq_loss}
 L( y_{\text{pred}}, y_{\text{true}})  = \frac{1}{N} \sum_{i=1}^{N} \left[ (y_{\text{true},i} - y_{\text{pred},i})^2 + (\beta - 1) \cdot \mathbf{1}_{(y_{\text{true},i} \geq \upsilon)} \cdot (y_{\text{true},i} - y_{\text{pred},i})^2 \right]
\end{equation}

\noindent
where $N$ is the batch size, $i$ is the index, $y_{\text{pred}}$ is the model output and $y_{\text{true}}$ is the true label.
  
\subsection{Few-shot training workflow for limited NCS data} 
We first measured the expression intensity of these 73 GFP NCS variants using fluorescence quantification. The sequences of these NCS and their corresponding fluorescence values formed our initial training dataset. For instance, the sequence C$_{4}$ showed an average fluorescence of 35,837 (standard deviation, $\sigma = 337.0$), while sequence Hag recorded 23,046 ($\sigma = 324.3$). Faced with the initial dataset that was both limited in size and imbalanced in distribution—with only one sequence exceeding a luminosity of 30,000, eight in the range of 20,000 to 30,000, 11 between 10,000 to 20,000, and 53 below 10,000—we adapted data augmentation in this study. Considering the static character of the genetic sequences, we focused on augmenting labels (the average fluorescence). We augmented each label by adding the product of Gaussian noise (with a mean of 1 and a variance of 0) with the standard deviation of the fluorescence for each NCS ~\cite{pukelsheim1994three}. We integrate these NCSs and their adjusted labels back into the original data to make the larger and more balanced.

Let datasets \( \{(x_i, y_i)\}_{i=1}^n \) be given ($n=73$ in the initial training dataset), where the true function \( f^* \) maps \( x \) to \( y \) as \( y_i = f^*(x_i) \), \( i = 1, \ldots, n \) (we use the biology verification to obtain this true function). 
We utilize a parameterized model \( f_{\theta}(x) \) to approximate $f^*$. 
We define a gene locus mutation function, $G(x)$, where each trinucleotide has a 20\% chance of being altered into a new, contextually suitable trinucleotide. 
Without loss of generality, we assume they are bounded: \( \sup_x f^*(x) < \infty \) and \( \sup_x f_{\theta}(x) < \infty \). 

Given the constrained availability of scientific data, traditional machine learning techniques may not be able to fit the target function \( f^* \) over the entire domain. Our algorithm focus on the training of \(x\) with the highest expression intensity and its corresponding \(G(x)\) using the following steps:

\begin{enumerate}
    \item Data augmentation;
    \item Train the model parameters: \( \theta_n = \arg\min_\theta \sum_{i=1}^n L(f_\theta(x_i), y_i) \);
    \item Directly searching for potential high-fluorescence NCSs \(\hat{\bf{x}} = G(\arg\max_{x} f_{\theta_{n}}(x))\);
    \item Greedily search for the top 10 NCSs predicted by our model: \\
    \mbox{}\hspace{0.5 cm}\(\{(\hat{x}_1, f_\theta(\hat{x}_1)), \ldots, (\hat{x}_{10}, f_\theta(\hat{x}_{10})\}\);
    \item Metabolic engineering verification and training dataset update: \\
    \mbox{}\hspace{0.5 cm}\(\{(x_i, y_i)\}_{i=1}^{n'} = \{(x_i, y_i)\}_{i=1}^{n} \cup \{\hat{\bf{x}}, f^* (\hat{\bf{x}}))\} \), \(n' = n+10\);
    \item Repeat step 1 to 5.
\end{enumerate}

Further, our approach focused on maximizing the quality of the solution using a single model in step 2 initially, de-emphasizing robustness. However, after identifying optimal genotypes, we then redirect our effort to enhancing the model's generalizability and robustness. We accomplished this by training different models with different \(beta\) mentioned in Equation ~\ref{eq_loss} before predicting the overall NCS expression intensity by a process of voting.

\subsection{NCS expression intensity analysis}  
\begin{figure}[ht]
    \centering
    \includegraphics[width = 1\linewidth]{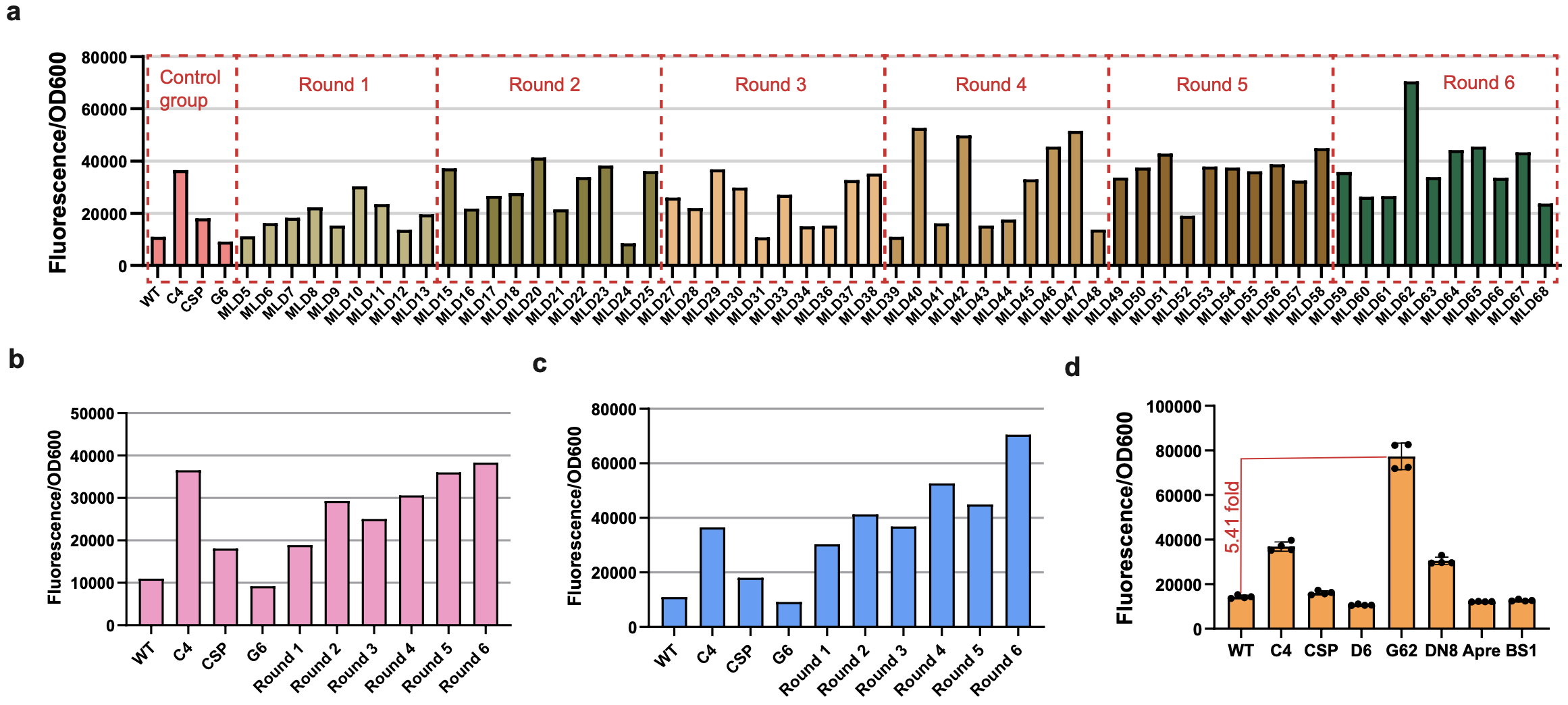}
        \caption{NCS expression intensity analysis.}
    \label{fig:subfig_g}
\begin{tablenotes}

\item {(a) Detailed fluorescence of each round;
(b) Average fluorescence of each round; (c) Max fluorescence of each round. The notation MLD$_{k}$ represents the machine learning-designed NCS with index \(k\). The endogenous NCS variants C$_{4}$, CSP, and G$_{6}$ serve as baselines, ordered from the largest to the smallest. The term WT refers to the baseline without the application of NCS for expression enhancement.}
\end{tablenotes}
\end{figure}

In Figure~\ref{fig:subfig_g}, we present the outcomes of six experimental cycles.  The initial cycle achieved an average fluorescence intensity of $18,901.1$, which was a 1.72-fold increase compared to the baseline \(WT\). The variant D10 showed a notable 2.76-fold increase but still ranked below the C$_{4}$ baseline with a fluorescence of 36,500. After this initial phase, we incorporated NCS variants from MLD$_{5}$ to MLD$_{13}$ into our training dataset, refining the model for future NCS predictions. Certain MLD$_{k}$ may be non-sequential because some NCS configurations were incompatible at the cellular level, and their fluorescence could not be measured. In the subsequent cycle, the average fluorescence rose to $29,271.6$, a 2.66-fold increase, surpassing the initial results. MLD$_{20}$ especially achieved a 3.76-fold rise, exceeding the highest-fluorescence C$_{4}$ in the original training dataset. The following cycles showed a consistent increase in average fluorescence: $25,040.2$ in the third, $30,602.0$ in the fourth, $36,032.1$ in the fifth, and reaching $38,316.6$ in the sixth. Among these, MLD$_{62}$ in the final cycle surpassed the C$_{4}$ reference, attaining a peak of $70,491$, almost double that of C$_{4}$, with the cycle's average fluorescence also exceeding the C$_{4}$ benchmark. We tried two more rounds but the result did not improve and so we stopped. 

Over the six iterative cycles, our model yielded NCSs that exhibited increased fluorescence, outperforming the commonly used endogenous NCS (C$_{4}$). 
We then performed repeated testing using our leading NCS design (MLD$_{62}$) as well as other state-of-the-art NCS designs, including DN8~\cite{tian2019synthetic}, BS1~\cite{wang2022model} and Apre~\cite{xu2021rational}, in order to verify correctness.

As shown in Figure~\ref{fig:subfig_g}(d) and Table~\ref{tabel:compare}, MLD$_{62}$ GFP expression increased by an average 5.41-fold compared to the $WT$ (wide type, implementation without NCS). We further compared our results against other leading NCS designs. Xu \textit{et al.}~\cite{xu2021rational} employed a statistical model to predict NCS-driven protein expression changes in {\em Bacillus subtilis WB600}, achieving a 0.85-fold enhancement using factors like G/C codon frequency and mRNA energy. Wang \textit{et al.}~\cite{wang2022model} utilized multi-view learning for synthetic NCS design in both {\em S. cerevisiae} and {\em B. subtilis}, attaining a 0.89-fold increase. Tian \textit{et al.}~\cite{tian2019synthetic} experimentally characterized 96 {\em B. subtilis} NCSs, observing a 2.13-fold gene expression enhancement. Comapred to these traditional, labor-intensive methods, our method is able to deliver better results using only six experimental cycles, requiring only 59 biology experiments in total.

\begin{table}[bt]
\scalebox{1.2}{

\begin{tabular}{c|ccc}
\hline
     & Methods                  & \begin{tabular}[c]{@{}c@{}}Number of \\ designed NCS\end{tabular} & \begin{tabular}[c]{@{}c@{}}Fold increase \\ of GFP expression\end{tabular} \\ \hline
DN8~\cite{tian2019synthetic}  & Statistics-guided                   & 1358                                                              & 2.13                                                                        \\
BS1~\cite{wang2022model}   & Model-driven                  & 5521                                                              & 0.89                                                                        \\
Apre~\cite{xu2021rational}  & Rational Design                   & 172                                                               & 0.85                                                                        \\ \hline
Ours(MLD$_{62}$)  & Few-shot learning  & \textbf{59}                                                       & \textbf{5.41}                                                                         \\ \hline
\end{tabular}

}
\caption{Comparison with state-of-art NCS designs.}
\label{tabel:compare}
\end{table}

\subsection{Computing resource}
We implemented our detection model using the CUDA-accelerated PyTorch versions 1.6.0 and 1.7.1. Specifically, the NCS detection model was trained on PyTorch version 1.7.1. All experiments were conducted on an Intel Xeon E5-2680 server, boasting 256GB DRAM and running 64-bit Linux 4.15.0. This server was equipped with both an Nvidia Tesla P100 GPU and a GeForce RT 3090 GPU.

\begin{figure*}[ht]
    \centering
    \includegraphics[width = 0.8\linewidth]{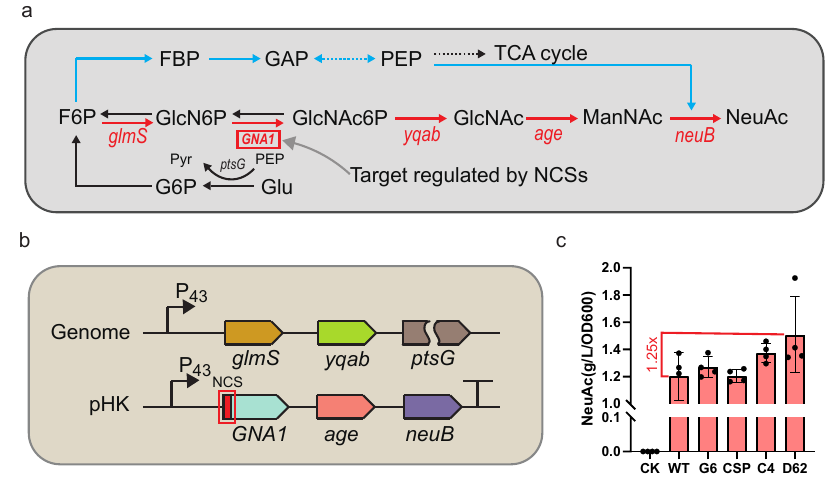}
        \caption{Efficient Synthesis of N-Acetylneuraminic Acid (NeuAc) using MLD-NCSs. (a) Synthetic pathway of NeuAc in Bacillus subtilis. The pathway involves several key genes (highlighted in red) and metabolic products including F6P (fructose-6-phosphate), GlcN6P (glucosamine-6-phosphate), GlcNAc6P (N-acetylglucosamine-6-phosphate), GlcNAc (N-acetylglucosamine), ManNAc (N-acetylmannosamine), and NeuAc (N-acetylneuraminic acid). Key genes in the pathway are glmS (glutamine-fructose-6-phosphate aminotransferase), GNA1 (glucosamine-6-phosphate N-acetyltransferase), yqaB (N-acetylglucosamine-6-phosphate phosphatase), age (N-acetylglucosamine-2-epimerase), and neuB (N-acetylneuraminic acid synthase), with ptsG (phosphotransferase) also involved. GNA1, the rate-limiting enzyme, was regulated by NCS. (b) NeuAc synthesis strategy in Bacillus subtilis. The genomic integration of glmS and yqaB genes, coupled with the deletion of ptsG and plasmid-based expression of GNA1, age, and neuB genes, is employed. GNA1, as the critical rate-limiting step, is targeted for regulation by NCS at its N-terminal. (c) Fermentation Results. The optimized MLD$_{62}$ variant showed a 25.3\% increase in NeuAc production compared to the wild type, and a 9.6\% increase over the most potent natural NCS variant.}
    \label{ecoll}
\end{figure*}

\subsection{Application of the MLD-NCSs}
To demonstrate the practical value of the MLD-NCSs, we targeted the critical rate-limiting gene GNA1 in NeuAc (N-acetylneuraminic acid) synthesis using MLD$_{62}$ regulation. Traditionally, without machine learning assistance, researchers have to construct extensive libraries and develop specialized screening strategies to identify potent regulatory elements. These elements, however, are often biologically constrained: overly strong regulation may not be conducive to cellular growth, preventing their selection. NeuAc, a high-value compound, has seen relative maturity in its biosynthetic production. However, a significant challenge remains in adequately regulating key rate-limiting steps. Using GNA1 as a target for NCS regulation, we found notable improvements. The GNA1 strain utilizing MLD$_{62}$ achieved a production intensity of 1.51g/L/OD600, which is 1.25-fold higher than the wild-type GNA1 (1.2g/L/OD600), and 1.1-fold higher than the strongest natural NCS-C4 (1.38g/L/OD600). This enhancement, particularly at the translational level, is significant as it substantially increases NeuAc yield without notably impacting normal cellular transcription. MLD$_{62}$ represents a powerful, MLD-based, non-natural NCS. Its superior regulatory capability, far exceeding that of natural NCS, was previously demonstrated using green fluorescent protein characterization, and this study further validates its value in practical applications.

\section{Discussion}\label{sec3}
We present a deep learning/synthetic biology co-designed workflow in optimizing the N-terminal coding sequence, a crucial factor influencing protein translation efficiency. By harnessing the synergy of deep learning, few-shot training, and metabolic modulation, we found a more efficient pathway for NCS refinement. We prove that our approach, integrating k-nearest encoding and word2vec algorithms for NCS encoding and utilizing attention mechanisms within a time-series prediction network, is effective in enhancing gene expression with limited training data. Using only six experimental iterations, we successfully engineered a NCS variant MLD$_{62}$ that outperforms all others reported so far by a significant margin. In the spirit of collaborative advancement and verifiability, we have made our GFP NCS expression database, the experiment protocol, and the model used accessible to the scientific community.

\section{Methods}\label{sec4}

\subsection{Strains, Plasmids, and Culturing Conditions}

In this study, we employed bacterial strains {\em B. subtilis} 168 and {\em E. coli} BL21(DE3) and utilized plasmid backbones pHK and pET28a. Strain descriptions are provided in Table ~\ref{dstrains}, and plasmid details are in Table ~\ref{plasmid}.


We used GFP to measure NCS intensity. Using the Gibson assembly method, we linked the GFP to the P43 promoter to generate the plasmid pHK-gfp harboring NCS. These were transformed into {\em B. subtilis} as a common platform for all NCS characterizations. Based on this strategy, 73 distinct NCSs were integrated into the N-terminal of the GFP through primers, producing the pHK-gfp plasmids, which were then individually transformed into {\em B. subtilis}.

For the training part of the algorithm, we measured fluorescence intensity by cultivating {\em B. subtilis} in 24-well deep plates. Each strain was initially inoculated into 1 mL LB medium (containing 5 g/L yeast extract, 10 g/L tryptone, and 10 g/L NaCl) and incubated overnight at 37°C with 220 rpm agitation. Afterward, \SI{10}{\micro\liter} of this seed culture was added to fresh LB medium and grown until saturation before recording fluorescence. To validate the training results, {\em B. subtilis} was cultivated under identical conditions, with four replicates per strain.

For {\em E. coli} BL21(DE3), we induced GFP expression in 24-well plates. Strains were inoculated into LB medium and incubated similarly overnight. \SI{100}{\micro\liter} seed culture was then introduced into fresh LB and grown for 2 hours before inducing with 2mmol IPTG. After 10 hours, fluorescence was quantified. Strains were preserved in glycerol at $-80$°C.

\begin{table}[hbp]
\begin{tabular}{c|c}
\hline
\textbf{Name} & \textbf{Relevant Characteristics}                                                                                        \\ \hline
BSU168        & {\em B. subtilis} 168 trpC2                                                                                                    \\ \hline
BSU168-comk   & \begin{tabular}[c]{@{}c@{}}Derivative of BSU168, expresses comK \\ gene under the control of P$_{\text{\em xylA}}$ promoter\end{tabular} \\ \hline
BSU168-MLDN   & \begin{tabular}[c]{@{}c@{}}Series derived from BSU168-comk,\\  each containing plasmid pHK-MLDN\end{tabular}             \\ \hline

{\em E. coli} JM109   & Commonly used plasmid construction and amplification hosts
           \\ \hline

{\em E. coli} BL21(DE3)   & A strain of {\em E. coli}         \\ \hline

\end{tabular}
\caption{Descriptions of strains.}
\label{dstrains}
\end{table}

\begin{table}[bt]
\begin{tabular}{c|c}
\hline
\textbf{Name}        & \textbf{Relevant Characteristics}                                                                                                                                                                     \\ \hline
\textbf{pHK}         & Kmr, {\em E. coli}-{\em B. subtilis} shuttle plasmid                                                                                                                                                        \\ \hline
\textbf{pHK-gfp}     & \begin{tabular}[c]{@{}c@{}}Derivative of pHK, expressing gfp \\ gene under P43 promoter\end{tabular}                                                                                                  \\ \hline
\textbf{pHK-Ngfp}    & \begin{tabular}[c]{@{}c@{}}73 series derived from pHK-gfp, each adding \\ different NCS sequences from Table 2-3 \\ before the ATG start codon at the 5’-end of gfp gene\end{tabular}                \\ \hline
\textbf{pHK-MLDNgfp} & \begin{tabular}[c]{@{}c@{}}Series derived from pHK-gfp,  each adding \\ a machine learning designed NCS sequence \\ identified as N before the ATG start codon at the 5’-end of gfp gene\end{tabular} \\ \hline

\textbf{pET28a}   &Km$^r$, a commonly used expression vector in Escherichia coli            \\ \hline

\textbf{pET28a-Ngfp}   & \begin{tabular}[c]{@{}c@{}}pET28a-derived plasmid, 
\\ expressing the gfp gene under the T7 promoter.\end{tabular}             \\ \hline

\end{tabular}
\caption{Descriptions of plasmid.}
\label{plasmid}
\end{table}

\subsection{Plasmid Construction}
All plasmids in this study are assembled using the Gibson Assembly method. Equal molar amounts of plasmid vector and PCR products are combined in the Gibson Assembly system and incubated at 50°C for one hour. This mixture is then transformed into {\em E. coli} cloning hosts. Plasmids that pass colony PCR and sequencing are further transformed into {\em B. subtilis}.

For the incorporation of NCS on the plasmid, all natural NCS from Table 2-3 are engineered into primers. A reverse PCR is conducted on the template, establishing an overlap region of approximately 15 to 20 base pairs. The resulting PCR products are directly transformed into the cloning host. Following successful sequencing verification, the obtained transformants yield the corresponding plasmids with NCS insertions.

\subsection{Strain Construction}

In this study, all strains harbor integrated P$_{\text{\em xylA}}$-comK expression cassettes in their genomes, allowing cells to be induced to express ComK by a xylose-inducible promoter, thereby facilitating competency. This feature streamlines the strain genome editing and plasmid transformation processes.

The procedure begins with inoculating a single colony into 3 mL of LB medium, followed by overnight incubation at 37°C and 220 rpm. The culture is then diluted five-fold and supplemented with a final concentration of 3\% xylose. After a subsequent 2-hour incubation at 37°C, the cells attain competency. For transformation, more than 100 ng of plasmid is introduced to the competent cells, which are then incubated at 37°C for over an hour. This step is followed by antibiotic selection on plates, utilizing kanamycin at a final concentration of \SI{50}{\gram\per\milli\liter} for {\em B. subtilis}.

In this study, {\em E. coli} competent cells were prepared using Sangon Biotech's (Shanghai) Super Competent Cell Preparation Kit.

For plasmid DNA chemical transformation: Super competent cells, stored at $-80^{\circ}$C, were thawed on ice. 100 ng of plasmid DNA (or 10 \(\mu\)L of cloned assembly product) was mixed with 100 \(\mu\)L of {\em E. coli} competent cells and left on ice for 30 minutes. The solution was heat-shocked at 42\(^{\circ}\)C for 45 seconds, then cooled on ice for 3 minutes. Subsequently, 750-1000 \(\mu\)L of antibiotic-free LB medium was added within a laminar flow hood. The mixture was then incubated at 37\(^{\circ}\)C, shaking at 220 RPM for 1 hour. After centrifuging the culture at 5000 RPM for 5 minutes, the supernatant was removed. The cells were then resuspended and spread onto LB agar plates with the corresponding antibiotic. Plates were cultured at 37\(^{\circ}\)C for 12-16 hours to identify positive transformants. The kanamycin concentration used for {\em E. coli} was set at 75 \(\mu\)g/mL.

\subsection{Analytical Method}

A Cytation 3 Multi-Mode Reader (BIOTEK) was used to measure the biomass and fluorescence intensity in each well of a 96-well plate containing \SI{200}{\micro\liter} of bacterial suspension. Biomass is characterized using OD600, defined here as the absorbance obtained at 600 nm in the reader for each well containing \SI{200}{\micro\liter} liquid in a 96-well plate. Fluorescence intensity is measured in arbitrary units, defined as the fluorescence intensity value measured in the reader at an excitation wavelength of 488 nm and emission wavelength of 523 nm for each well containing  \SI{200}{\micro\liter}  liquid in a 96-well plate.

\newpage
\bibliography{sn-bibliography}


\begin{thebibliography}{25}
\ifx \bisbn   \undefined \def \bisbn  #1{ISBN #1}\fi
\ifx \binits  \undefined \def \binits#1{#1}\fi
\ifx \bauthor  \undefined \def \bauthor#1{#1}\fi
\ifx \batitle  \undefined \def \batitle#1{#1}\fi
\ifx \bjtitle  \undefined \def \bjtitle#1{#1}\fi
\ifx \bvolume  \undefined \def \bvolume#1{\textbf{#1}}\fi
\ifx \byear  \undefined \def \byear#1{#1}\fi
\ifx \bissue  \undefined \def \bissue#1{#1}\fi
\ifx \bfpage  \undefined \def \bfpage#1{#1}\fi
\ifx \blpage  \undefined \def \blpage #1{#1}\fi
\ifx \burl  \undefined \def \burl#1{\textsf{#1}}\fi
\ifx \doiurl  \undefined \def \doiurl#1{\url{https://doi.org/#1}}\fi
\ifx \betal  \undefined \def \betal{\textit{et al.}}\fi
\ifx \binstitute  \undefined \def \binstitute#1{#1}\fi
\ifx \binstitutionaled  \undefined \def \binstitutionaled#1{#1}\fi
\ifx \bctitle  \undefined \def \bctitle#1{#1}\fi
\ifx \beditor  \undefined \def \beditor#1{#1}\fi
\ifx \bpublisher  \undefined \def \bpublisher#1{#1}\fi
\ifx \bbtitle  \undefined \def \bbtitle#1{#1}\fi
\ifx \bedition  \undefined \def \bedition#1{#1}\fi
\ifx \bseriesno  \undefined \def \bseriesno#1{#1}\fi
\ifx \blocation  \undefined \def \blocation#1{#1}\fi
\ifx \bsertitle  \undefined \def \bsertitle#1{#1}\fi
\ifx \bsnm \undefined \def \bsnm#1{#1}\fi
\ifx \bsuffix \undefined \def \bsuffix#1{#1}\fi
\ifx \bparticle \undefined \def \bparticle#1{#1}\fi
\ifx \barticle \undefined \def \barticle#1{#1}\fi
\bibcommenthead
\ifx \bconfdate \undefined \def \bconfdate #1{#1}\fi
\ifx \botherref \undefined \def \botherref #1{#1}\fi
\ifx \url \undefined \def \url#1{\textsf{#1}}\fi
\ifx \bchapter \undefined \def \bchapter#1{#1}\fi
\ifx \bbook \undefined \def \bbook#1{#1}\fi
\ifx \bcomment \undefined \def \bcomment#1{#1}\fi
\ifx \oauthor \undefined \def \oauthor#1{#1}\fi
\ifx \citeauthoryear \undefined \def \citeauthoryear#1{#1}\fi
\ifx \endbibitem  \undefined \def \endbibitem {}\fi
\ifx \bconflocation  \undefined \def \bconflocation#1{#1}\fi
\ifx \arxivurl  \undefined \def \arxivurl#1{\textsf{#1}}\fi
\csname PreBibitemsHook\endcsname

\bibitem[\protect\citeauthoryear{Horton et~al.}{2023}]{horton2023short}
\begin{barticle}
\bauthor{\bsnm{Horton}, \binits{C.A.}},
\bauthor{\bsnm{Alexandari}, \binits{A.M.}},
\bauthor{\bsnm{Hayes}, \binits{M.G.}},
\bauthor{\bsnm{Marklund}, \binits{E.}},
\bauthor{\bsnm{Schaepe}, \binits{J.M.}},
\bauthor{\bsnm{Aditham}, \binits{A.K.}},
\bauthor{\bsnm{Shah}, \binits{N.}},
\bauthor{\bsnm{Suzuki}, \binits{P.H.}},
\bauthor{\bsnm{Shrikumar}, \binits{A.}},
\bauthor{\bsnm{Afek}, \binits{A.}}, \betal:
\batitle{Short tandem repeats bind transcription factors to tune eukaryotic
  gene expression}.
\bjtitle{Science}
\bvolume{381}(\bissue{6664}),
\bfpage{1250}
(\byear{2023})
\end{barticle}
\endbibitem

\bibitem[\protect\citeauthoryear{Gil and Ulitsky}{2020}]{gil2020regulation}
\begin{barticle}
\bauthor{\bsnm{Gil}, \binits{N.}},
\bauthor{\bsnm{Ulitsky}, \binits{I.}}:
\batitle{Regulation of gene expression by cis-acting long non-coding rnas}.
\bjtitle{Nature Reviews Genetics}
\bvolume{21}(\bissue{2}),
\bfpage{102}--\blpage{117}
(\byear{2020})
\end{barticle}
\endbibitem

\bibitem[\protect\citeauthoryear{Bosch et~al.}{2021}]{bosch2021genome}
\begin{barticle}
\bauthor{\bsnm{Bosch}, \binits{B.}},
\bauthor{\bsnm{DeJesus}, \binits{M.A.}},
\bauthor{\bsnm{Poulton}, \binits{N.C.}},
\bauthor{\bsnm{Zhang}, \binits{W.}},
\bauthor{\bsnm{Engelhart}, \binits{C.A.}},
\bauthor{\bsnm{Zaveri}, \binits{A.}},
\bauthor{\bsnm{Lavalette}, \binits{S.}},
\bauthor{\bsnm{Ruecker}, \binits{N.}},
\bauthor{\bsnm{Trujillo}, \binits{C.}},
\bauthor{\bsnm{Wallach}, \binits{J.B.}}, \betal:
\batitle{Genome-wide gene expression tuning reveals diverse vulnerabilities of
  m. tuberculosis}.
\bjtitle{Cell}
\bvolume{184}(\bissue{17}),
\bfpage{4579}--\blpage{4592}
(\byear{2021})
\end{barticle}
\endbibitem

\bibitem[\protect\citeauthoryear{Fu et~al.}{2022}]{fu2022operator}
\begin{barticle}
\bauthor{\bsnm{Fu}, \binits{G.}},
\bauthor{\bsnm{Yue}, \binits{J.}},
\bauthor{\bsnm{Li}, \binits{D.}},
\bauthor{\bsnm{Li}, \binits{Y.}},
\bauthor{\bsnm{Lee}, \binits{S.Y.}},
\bauthor{\bsnm{Zhang}, \binits{D.}}:
\batitle{An operator-based expression toolkit for bacillus subtilis enables
  fine-tuning of gene expression and biosynthetic pathway regulation}.
\bjtitle{Proceedings of the National Academy of Sciences}
\bvolume{119}(\bissue{11}),
\bfpage{2119980119}
(\byear{2022})
\end{barticle}
\endbibitem

\bibitem[\protect\citeauthoryear{Lu et~al.}{2019}]{lu2019crispr}
\begin{barticle}
\bauthor{\bsnm{Lu}, \binits{Z.}},
\bauthor{\bsnm{Yang}, \binits{S.}},
\bauthor{\bsnm{Yuan}, \binits{X.}},
\bauthor{\bsnm{Shi}, \binits{Y.}},
\bauthor{\bsnm{Ouyang}, \binits{L.}},
\bauthor{\bsnm{Jiang}, \binits{S.}},
\bauthor{\bsnm{Yi}, \binits{L.}},
\bauthor{\bsnm{Zhang}, \binits{G.}}:
\batitle{Crispr-assisted multi-dimensional regulation for fine-tuning gene
  expression in bacillus subtilis}.
\bjtitle{Nucleic acids research}
\bvolume{47}(\bissue{7}),
\bfpage{40}--\blpage{40}
(\byear{2019})
\end{barticle}
\endbibitem

\bibitem[\protect\citeauthoryear{Ding et~al.}{2020}]{ding2020programmable}
\begin{barticle}
\bauthor{\bsnm{Ding}, \binits{N.}},
\bauthor{\bsnm{Yuan}, \binits{Z.}},
\bauthor{\bsnm{Zhang}, \binits{X.}},
\bauthor{\bsnm{Chen}, \binits{J.}},
\bauthor{\bsnm{Zhou}, \binits{S.}},
\bauthor{\bsnm{Deng}, \binits{Y.}}:
\batitle{Programmable cross-ribosome-binding sites to fine-tune the dynamic
  range of transcription factor-based biosensor}.
\bjtitle{Nucleic Acids Research}
\bvolume{48}(\bissue{18}),
\bfpage{10602}--\blpage{10613}
(\byear{2020})
\end{barticle}
\endbibitem

\bibitem[\protect\citeauthoryear{Lv et~al.}{2023}]{lv2023crispr}
\begin{botherref}
\oauthor{\bsnm{Lv}, \binits{X.}},
\oauthor{\bsnm{Li}, \binits{Y.}},
\oauthor{\bsnm{Xiu}, \binits{X.}},
\oauthor{\bsnm{Liao}, \binits{C.}},
\oauthor{\bsnm{Xu}, \binits{Y.}},
\oauthor{\bsnm{Liu}, \binits{Y.}},
\oauthor{\bsnm{Li}, \binits{J.}},
\oauthor{\bsnm{Du}, \binits{G.}},
\oauthor{\bsnm{Liu}, \binits{L.}}:
Crispr genetic toolkits of classical food microorganisms: Current state and
  future prospects.
Biotechnology Advances,
108261
(2023)
\end{botherref}
\endbibitem

\bibitem[\protect\citeauthoryear{Yang et~al.}{2017}]{yang2017characterization}
\begin{barticle}
\bauthor{\bsnm{Yang}, \binits{S.}},
\bauthor{\bsnm{Du}, \binits{G.}},
\bauthor{\bsnm{Chen}, \binits{J.}},
\bauthor{\bsnm{Kang}, \binits{Z.}}:
\batitle{Characterization and application of endogenous phase-dependent
  promoters in bacillus subtilis}.
\bjtitle{Applied microbiology and biotechnology}
\bvolume{101},
\bfpage{4151}--\blpage{4161}
(\byear{2017})
\end{barticle}
\endbibitem

\bibitem[\protect\citeauthoryear{Tian et~al.}{2019}]{tian2019synthetic}
\begin{barticle}
\bauthor{\bsnm{Tian}, \binits{R.}},
\bauthor{\bsnm{Liu}, \binits{Y.}},
\bauthor{\bsnm{Chen}, \binits{J.}},
\bauthor{\bsnm{Li}, \binits{J.}},
\bauthor{\bsnm{Liu}, \binits{L.}},
\bauthor{\bsnm{Du}, \binits{G.}},
\bauthor{\bsnm{Chen}, \binits{J.}}:
\batitle{Synthetic n-terminal coding sequences for fine-tuning gene expression
  and metabolic engineering in bacillus subtilis}.
\bjtitle{Metabolic engineering}
\bvolume{55},
\bfpage{131}--\blpage{141}
(\byear{2019})
\end{barticle}
\endbibitem

\bibitem[\protect\citeauthoryear{Fredrick and
  Ibba}{2010}]{fredrick2010sequence}
\begin{barticle}
\bauthor{\bsnm{Fredrick}, \binits{K.}},
\bauthor{\bsnm{Ibba}, \binits{M.}}:
\batitle{How the sequence of a gene can tune its translation}.
\bjtitle{Cell}
\bvolume{141}(\bissue{2}),
\bfpage{227}--\blpage{229}
(\byear{2010})
\end{barticle}
\endbibitem

\bibitem[\protect\citeauthoryear{Tian et~al.}{2020}]{tian2020titrating}
\begin{barticle}
\bauthor{\bsnm{Tian}, \binits{R.}},
\bauthor{\bsnm{Liu}, \binits{Y.}},
\bauthor{\bsnm{Cao}, \binits{Y.}},
\bauthor{\bsnm{Zhang}, \binits{Z.}},
\bauthor{\bsnm{Li}, \binits{J.}},
\bauthor{\bsnm{Liu}, \binits{L.}},
\bauthor{\bsnm{Du}, \binits{G.}},
\bauthor{\bsnm{Chen}, \binits{J.}}:
\batitle{Titrating bacterial growth and chemical biosynthesis for efficient
  n-acetylglucosamine and n-acetylneuraminic acid bioproduction}.
\bjtitle{Nature Communications}
\bvolume{11}(\bissue{1}),
\bfpage{5078}
(\byear{2020})
\end{barticle}
\endbibitem

\bibitem[\protect\citeauthoryear{Zhao et~al.}{2021}]{zhao2021directed}
\begin{barticle}
\bauthor{\bsnm{Zhao}, \binits{H.}},
\bauthor{\bsnm{Ding}, \binits{W.}},
\bauthor{\bsnm{Zang}, \binits{J.}},
\bauthor{\bsnm{Yang}, \binits{Y.}},
\bauthor{\bsnm{Liu}, \binits{C.}},
\bauthor{\bsnm{Hu}, \binits{L.}},
\bauthor{\bsnm{Chen}, \binits{Y.}},
\bauthor{\bsnm{Liu}, \binits{G.}},
\bauthor{\bsnm{Fang}, \binits{Y.}},
\bauthor{\bsnm{Yuan}, \binits{Y.}}, \betal:
\batitle{Directed-evolution of translation system for efficient unnatural amino
  acids incorporation and generalizable synthetic auxotroph construction}.
\bjtitle{Nature Communications}
\bvolume{12}(\bissue{1}),
\bfpage{7039}
(\byear{2021})
\end{barticle}
\endbibitem

\bibitem[\protect\citeauthoryear{Stork et~al.}{2021}]{stork2021designing}
\begin{barticle}
\bauthor{\bsnm{Stork}, \binits{D.A.}},
\bauthor{\bsnm{Squyres}, \binits{G.R.}},
\bauthor{\bsnm{Kuru}, \binits{E.}},
\bauthor{\bsnm{Gromek}, \binits{K.A.}},
\bauthor{\bsnm{Rittichier}, \binits{J.}},
\bauthor{\bsnm{Jog}, \binits{A.}},
\bauthor{\bsnm{Burton}, \binits{B.M.}},
\bauthor{\bsnm{Church}, \binits{G.M.}},
\bauthor{\bsnm{Garner}, \binits{E.C.}},
\bauthor{\bsnm{Kunjapur}, \binits{A.M.}}:
\batitle{Designing efficient genetic code expansion in bacillus subtilis to
  gain biological insights}.
\bjtitle{Nature Communications}
\bvolume{12}(\bissue{1}),
\bfpage{5429}
(\byear{2021})
\end{barticle}
\endbibitem

\bibitem[\protect\citeauthoryear{Cambray et~al.}{2018}]{cambray2018evaluation}
\begin{barticle}
\bauthor{\bsnm{Cambray}, \binits{G.}},
\bauthor{\bsnm{Guimaraes}, \binits{J.C.}},
\bauthor{\bsnm{Arkin}, \binits{A.P.}}:
\batitle{Evaluation of 244,000 synthetic sequences reveals design principles to
  optimize translation in escherichia coli}.
\bjtitle{Nature biotechnology}
\bvolume{36}(\bissue{10}),
\bfpage{1005}--\blpage{1015}
(\byear{2018})
\end{barticle}
\endbibitem

\bibitem[\protect\citeauthoryear{Goodman et~al.}{2013}]{goodman2013causes}
\begin{barticle}
\bauthor{\bsnm{Goodman}, \binits{D.B.}},
\bauthor{\bsnm{Church}, \binits{G.M.}},
\bauthor{\bsnm{Kosuri}, \binits{S.}}:
\batitle{Causes and effects of n-terminal codon bias in bacterial genes}.
\bjtitle{Science}
\bvolume{342}(\bissue{6157}),
\bfpage{475}--\blpage{479}
(\byear{2013})
\end{barticle}
\endbibitem

\bibitem[\protect\citeauthoryear{Kudla et~al.}{2009}]{kudla2009coding}
\begin{barticle}
\bauthor{\bsnm{Kudla}, \binits{G.}},
\bauthor{\bsnm{Murray}, \binits{A.W.}},
\bauthor{\bsnm{Tollervey}, \binits{D.}},
\bauthor{\bsnm{Plotkin}, \binits{J.B.}}:
\batitle{Coding-sequence determinants of gene expression in escherichia coli}.
\bjtitle{science}
\bvolume{324}(\bissue{5924}),
\bfpage{255}--\blpage{258}
(\byear{2009})
\end{barticle}
\endbibitem

\bibitem[\protect\citeauthoryear{Espah~Borujeni
  et~al.}{2017}]{espah2017precise}
\begin{barticle}
\bauthor{\bsnm{Espah~Borujeni}, \binits{A.}},
\bauthor{\bsnm{Cetnar}, \binits{D.}},
\bauthor{\bsnm{Farasat}, \binits{I.}},
\bauthor{\bsnm{Smith}, \binits{A.}},
\bauthor{\bsnm{Lundgren}, \binits{N.}},
\bauthor{\bsnm{Salis}, \binits{H.M.}}:
\batitle{Precise quantification of translation inhibition by mrna structures
  that overlap with the ribosomal footprint in n-terminal coding sequences}.
\bjtitle{Nucleic acids research}
\bvolume{45}(\bissue{9}),
\bfpage{5437}--\blpage{5448}
(\byear{2017})
\end{barticle}
\endbibitem

\bibitem[\protect\citeauthoryear{Wang et~al.}{2022}]{wang2022model}
\begin{barticle}
\bauthor{\bsnm{Wang}, \binits{C.}},
\bauthor{\bsnm{Zhang}, \binits{W.}},
\bauthor{\bsnm{Tian}, \binits{R.}},
\bauthor{\bsnm{Zhang}, \binits{J.}},
\bauthor{\bsnm{Zhang}, \binits{L.}},
\bauthor{\bsnm{Deng}, \binits{Z.}},
\bauthor{\bsnm{Lv}, \binits{X.}},
\bauthor{\bsnm{Li}, \binits{J.}},
\bauthor{\bsnm{Liu}, \binits{L.}},
\bauthor{\bsnm{Du}, \binits{G.}}, \betal:
\batitle{Model-driven design of synthetic n-terminal coding sequences for
  regulating gene expression in yeast and bacteria}.
\bjtitle{Biotechnology Journal}
\bvolume{17}(\bissue{5}),
\bfpage{2100655}
(\byear{2022})
\end{barticle}
\endbibitem

\bibitem[\protect\citeauthoryear{Rong}{2014}]{rong2014word2vec}
\begin{botherref}
\oauthor{\bsnm{Rong}, \binits{X.}}:
word2vec parameter learning explained.
arXiv preprint arXiv:1411.2738
(2014)
\end{botherref}
\endbibitem

\bibitem[\protect\citeauthoryear{Vaswani et~al.}{2017}]{vaswani2017attention}
\begin{botherref}
\oauthor{\bsnm{Vaswani}, \binits{A.}},
\oauthor{\bsnm{Shazeer}, \binits{N.}},
\oauthor{\bsnm{Parmar}, \binits{N.}},
\oauthor{\bsnm{Uszkoreit}, \binits{J.}},
\oauthor{\bsnm{Jones}, \binits{L.}},
\oauthor{\bsnm{Gomez}, \binits{A.N.}},
\oauthor{\bsnm{Kaiser}, \binits{{\L}.}},
\oauthor{\bsnm{Polosukhin}, \binits{I.}}:
Attention is all you need.
Advances in neural information processing systems
\textbf{30}
(2017)
\end{botherref}
\endbibitem

\bibitem[\protect\citeauthoryear{Graves and Graves}{2012}]{graves2012long}
\begin{botherref}
\oauthor{\bsnm{Graves}, \binits{A.}},
\oauthor{\bsnm{Graves}, \binits{A.}}:
Long short-term memory.
Supervised sequence labelling with recurrent neural networks,
37--45
(2012)
\end{botherref}
\endbibitem

\bibitem[\protect\citeauthoryear{Liu et~al.}{2020}]{liu2020chassis}
\begin{barticle}
\bauthor{\bsnm{Liu}, \binits{J.}},
\bauthor{\bsnm{Wu}, \binits{X.}},
\bauthor{\bsnm{Yao}, \binits{M.}},
\bauthor{\bsnm{Xiao}, \binits{W.}},
\bauthor{\bsnm{Zha}, \binits{J.}}:
\batitle{Chassis engineering for microbial production of chemicals: from
  natural microbes to synthetic organisms}.
\bjtitle{Current Opinion in Biotechnology}
\bvolume{66},
\bfpage{105}--\blpage{112}
(\byear{2020})
\end{barticle}
\endbibitem

\bibitem[\protect\citeauthoryear{Gu et~al.}{2018}]{gu2018advances}
\begin{barticle}
\bauthor{\bsnm{Gu}, \binits{Y.}},
\bauthor{\bsnm{Xu}, \binits{X.}},
\bauthor{\bsnm{Wu}, \binits{Y.}},
\bauthor{\bsnm{Niu}, \binits{T.}},
\bauthor{\bsnm{Liu}, \binits{Y.}},
\bauthor{\bsnm{Li}, \binits{J.}},
\bauthor{\bsnm{Du}, \binits{G.}},
\bauthor{\bsnm{Liu}, \binits{L.}}:
\batitle{Advances and prospects of bacillus subtilis cellular factories: from
  rational design to industrial applications}.
\bjtitle{Metabolic engineering}
\bvolume{50},
\bfpage{109}--\blpage{121}
(\byear{2018})
\end{barticle}
\endbibitem

\bibitem[\protect\citeauthoryear{Xu et~al.}{2021}]{xu2021rational}
\begin{barticle}
\bauthor{\bsnm{Xu}, \binits{K.}},
\bauthor{\bsnm{Tong}, \binits{Y.}},
\bauthor{\bsnm{Li}, \binits{Y.}},
\bauthor{\bsnm{Tao}, \binits{J.}},
\bauthor{\bsnm{Li}, \binits{J.}},
\bauthor{\bsnm{Zhou}, \binits{J.}},
\bauthor{\bsnm{Liu}, \binits{S.}}:
\batitle{Rational design of the n-terminal coding sequence for regulating
  enzyme expression in bacillus subtilis}.
\bjtitle{ACS Synthetic Biology}
\bvolume{10}(\bissue{2}),
\bfpage{265}--\blpage{276}
(\byear{2021})
\end{barticle}
\endbibitem

\bibitem[\protect\citeauthoryear{Pukelsheim}{1994}]{pukelsheim1994three}
\begin{barticle}
\bauthor{\bsnm{Pukelsheim}, \binits{F.}}:
\batitle{The three sigma rule}.
\bjtitle{The American Statistician}
\bvolume{48}(\bissue{2}),
\bfpage{88}--\blpage{91}
(\byear{1994})
\end{barticle}
\endbibitem

\end{thebibliography}



\newpage
\appendix
\section{Appendix}
\subsection{Opensource dataset}
\begin{longtable}{llc}
\hline
Name  & NCS                                            & Fluorescence \\ \hline
$C_4$    & ATGAAAAAAATCACAACAAACGAACAATTTAATGAACTGATTCAA  & 35837        \\
E10   & ATGGAAATGATGATTAAAAAAAGAATTAAACAAGTCAAAAAAGGC  & 23510        \\
Hag   & ATGAGAATTAACCACAATATTGCAGCGCTTAACACACTGAACCGT  & 23046        \\
E6    & ATGTTTATGAAATCTACTGGTATTGTACGTAAAGTTGATGAATTA  & 22952        \\
B3    & ATGAATATAAATGTTGATGTGAAGCAAAACGAGAATGATATACAA  & 21946        \\
E4    & ATGAACTATAACATCAGAGGAGAAAATATTGAAGTGACACCCGCG  & 21802        \\
GlnA  & ATGGCAAAGTACACTAGAGAAGATATCGAAAAATTAGTAAAAGAA  & 20912        \\
IlvC  & ATGGTAAAAGTATATTATAACGGTGATATCAAAGAGAACGTATTG  & 20798        \\
CspD  & ATGCAAAACGGTAAAGTAAAATGGTTCAACAACGAAAAAGGATTC  & 20088        \\
E7    & ATGGAAACGAATGAACAAACAATGCCGACGAAATATGATCCGGCA  & 17889        \\
TufA  & ATGGCTAAAGAAAAATTCGACCGTTCCAAATCACATGCCAATATT  & 17256        \\
H2    & ATGATTACGAAAACTAGCAAAAATGCTGCTCGTCTTAAAAGACAC  & 15325        \\
A9    & ATGTCAGAACAGAAAAAAGTCGTATTAGCATACTCAGGAGGTCTT  & 14258        \\
E11   & ATGAGAAACGAACGCAGAAAAAAGAAAAAAACTTTATTACTGACA  & 13613        \\
E2    & ATGAGTATAAACATAAAAGCAGTAACTGATGATAATCGTGCTGCA  & 13603        \\
C10   & ATGTTTAAGCACACAAAAATGCTGCAGCATCCTGCTAAACCAGAT  & 13126        \\
A7    & ATGGCAGACAATAACAAAATGAGCAGAGAAGAAGCAGGTAGAAAA  & 12150        \\
E1    & ATGGCTGAATGGAAAACAAAACGGACATACGATGAGATATTGTAT  & 11568        \\
CspB  & ATGTTAGAAGGTAAAGTAAAATGGTTCAACTCTGAAAAAGGTTTC  & 10566        \\
F6    & ATGTCATTAAGAGAAGAAGCATTACACCTGCATAAAGTCAACCAG  & 10072        \\
B8    & ATGGCTCAACAAACGAATGTTGCAGGACAAAAAACAGAAAAACAA  & 9963         \\
A3    & ATGTCTGATTCAAATCTTACGAATCCTATAAAAGCATTTTTTCAT  & 9869         \\
E8    & ATGAGGAAAACAGTCATTGTAAGTGCTGCAAGAACTCCATTTGGC  & 9797         \\
G7    & ATGAGAAGCTATGAAAAATCAAAAACGGCTTTTAAAGAAGCGCAA  & 9163         \\
H3    & TTGAATCAAAAAGCTGTCATTCTCGACGAACAGGCAATTAGACGG  & 9024         \\
H9    & ATGGCCAAAATAAAAGATGATTGTATAGAACTTGAATTAACACCG  & 8810         \\
G6    & ATGACCATTAAACGTGCATTAATCAGTGTTTCAGATAAAACAAAT  & 7789         \\
H8    & ATGGCAGACACATTAGAGCGTGTAACGAAAATCATCGTAGATCGC  & 7147         \\
B4    & ATGGCACTATTTACAGCAAAAGTAACCGCGCGAGGCGGACGAGCA  & 6699         \\
B2    & ATGGGACTTTTAGAAGATTTGCAAAGACAGGTGTTAATCGGTGAC  & 6474         \\
10    & ATGGCAGGATTAATTCGTGTCACACCCGAAGAGCTAAGAGCGATG  & 6432         \\
F3    & ATGACTAAACAAACAATTCGCGTTGAATTGACATCAACAAAAAAA  & 6361         \\
A4    & ATGACCCATTCATTTGCTGTTCCACGTTCTGTTGAATGGAAAGAA  & 5987         \\
C6    & ATGGAACCTTTGAAATCACATACGGGGAAAGCAGCCGTATTAAAT  & 5914         \\
F2    & ATGAAAAAAATTCCGGTTACCGTACTGAGCGGTTATCTCGGTGCG  & 5615         \\
D1    & ATGTGCCAATCCAATCAAATTGTCAGCCATTTTTTATCCCATCGA  & 5392         \\
8     & ATGTCTTTAATCGGTAAAGAAGTACTTCCATTCGAAGCAAAAGCA  & 5187         \\
D7    & ATGGCTGCAAAACAAGAACGCTGGCGAGAGCTCGCTGAAGTAAAA  & 4906         \\
G3    & ATGTCGTTTTTCAGAAATCAATTAGCGAATGTAGTAGAGTGGGAA  & 4837         \\
G10   & ATGTTTCAAAATAGTATGAAACAACGAATGAATTGGGAAGATTTT  & 4690         \\
F9    & ATGGCAGCAAAATTTGAAGTGGGCAGTGTTTACACTGGTAAAGTT  & 4485         \\
D8    & ATGGTGACCAAAATTCTAAAAGCACCGGACGGCTCTCCAAGTGAT  & 4447         \\
H11   & ATGACCAAAGGAATCTTAGGAAGAAAAATTGGTATGACGCAAGTA  & 4438         \\
A5    & ATGACAACCATCAAAACATCGAATTTAGGATTTCCGAGAATCGGA  & 3894         \\
G4    & TTGATGTCGAACCAGACTGTATACCAGTTCATTGCCGAAAATCAA  & 3659         \\
2     & ATGGCTTTAAATATCGAAGAAATCATTGCTTCCGTTAAAGAAGCA  & 3364         \\
G9    & ATGAGAATGCGCCACAAGCCTTGGGCTGATGACTTTTTGGCTGAA  & 3260         \\
B1    & TTGAGGAAAGATGAAATCATGCATATCGTATCATGCGCAGATGAT  & 3026         \\
11    & ATGGATGCGCTTATTGAGGAAGTTGATGGCATTTCAAATCGTACT  & 2825         \\
B6    & ATGGCACATAGAATTTTAATTGTAGATGACGCAGCATTTATGCGA  & 2792         \\
C1    & ATGGGTCTTATTGTACAAAAATTCGGAGGCACTTCCGTCGGCTCA  & 2791         \\
A6    & ATGAGCAGCTTGTTTCAAACCTACGGCCGTTGGGATATTGACATC  & 2722         \\
C7    & ATGGCAAAAGTATTATATATCACTGCTCATCCACATGACGAAGCA  & 2622         \\
H12   & ATGATTATCTGTAAAACCCCACGTGAACTTGGTATCATGCGGGAA  & 2602         \\
H7    & ATGAAACGAGATAAGGTGCAGACCTTACATGGAGAAATACATATT  & 2503         \\
F4    & ATGTCTATGCATAAAGCACTCACCATTGCCGGCTCAGATTCCAGC  & 2422         \\
G8    & ATGGTGACAACGGTGCAGCGTACGTTCCGAAAGGAAGTTCTACAT  & 2340         \\
C5    & TTGAAGAAACGTATTGCTCTATTGCCCGGAGACGGGATCGGCCCT  & 2323         \\
D6    & ATGAACGACCAATCCTGTGTAAGAATCATGACAGAATGGGATATT  & 2260         \\
Icd   & GTGGCACAAGGTGAAAAAATTACAGTCTCTAACGGAGTATTAAAC  & 2128         \\
G5    & ATGATACGAAGTATGACAGGCTTCGGCAGTGCAAGCAAAACACAA  & 2015         \\
D4    & GTGACAAATCGCGATATTGTATGGCATGAAGCCTCTATCACAAAA  & 1943         \\
E5    & TTGTTATTTAAAAAAGACAGAAAACAAGAAACAGCTTACTTTTCA  & 1909         \\
B7    & TTGAAAATAGGAATTGTAGGTGCTACAGGATATGGAGGCACCGAA  & 1853         \\
C9    & TTGAGTAAACACAATTGGACGCTGGAAACCCAGCTCGTGCACAAT  & 1823         \\
F8    & GTGAAGTTTTCAGAAGAATGCCGCAGTGCAGCCGCAGAATGGTGG  & 1749         \\
D3    & ATGTACATATTTCAAGCTGATCAGCTTAGTGCCAAAGACACATAC  & 1691         \\
E9    & GTGAAAAATAAATGGCTGTCTTTTTTTTCGGGTAAGGTCCAGCTT  & 1389         \\
B10   & ATGAAAACAGACTGGTGGAAGGATGCAGTGGTGTACCAAATTTAC  & 1188         \\
9     & ATGAGAAAGTACGAAGTTATGTACATTATCCGCCCAAACATTGAC  & 1086         \\
E3    & GTGGAAGTTACTGACGTAAGATTACGCCGCGTGAATACCGATGGT  & 1056         \\
G2    & ATGGCGCAAATGACAATGATTCAAGCAATCACGGATGCGTTACGC  & 879          \\
H4    & ATGGAAAAAAAACCGTTAACTCCTAGACAGATTGTAGATCGGTTA  & 401          \\ 
MLD5  & ATGAAAAAAATCAGTAACAATGGACCAATAAACACAGTGATTCTC  & 11168        \\
MLD6  & ATGAAAAAAATGACGGTAAAGGCGGCTAAAAATACCAAGATCGCA  & 16221        \\
MLD7  & ATGAAAAAAAAAACCCAAAACGACGTAACAAATACGCTGAAACTG  & 18189        \\
MLD8  & ATGAAAAAGAACAGTTATAAGCGTGCGACAAAGACAACGAACGCC. & 22251        \\
MLD9  & ATGCGAAAGATAAGCCGAAATCGTGCCGAAAAAGAGAAGATCGCT. & 15253        \\
MLD10 & ATGAAAAAAATCACACGTAACGAACAATTTAATACGAAGATTCAA. & 30310        \\
MLD11 & ATGAAAAAAATCACAACAAACGAACAAACTAATACGAAGATTCAA  & 23499        \\
MLD12 & ATGAGAAAGATCACAACAAACCGCCAATTTAATGAACTGATTCAA  & 13662        \\
MLD13 & ATGAAAAAAATCACAACAAACGAACAAACAAATACGCTGATTCAA  & 19558        \\
MLD15 & ATGAAAAAAATCAGCACAAACGAAAATTTTAATGAACTGATTCAA  & 37184        \\
MLD16 & ATGAAAAAAATCACGACAAACATTCAAAACAATGAACTGATTCAA  & 21753        \\
MLD17 & ATGAAAAAAATCAGCACAAACGAACAAAAGAATGAACTGATTGGG  & 26650        \\
MLD18 & ATGAAAAAAATCAGTACAAACAGACAAAAGAATGAACTGATTCAA  & 27677        \\
MLD20 & ATGAAAAAAATCACAACAAACAGGCAAAACAATGAACTGAAACAA  & 41272        \\
MLD21 & ATGAAAAAAATCACGACAAACATTCAAGAGAATGAACTGATTCAA  & 21514        \\
MLD22 & ATGAAAAAAATCTCGACAAAAATGCAAAACAATGAACTGATTCAA  & 33849        \\
MLD23 & ATGAAAAAATACAGCACAAACAGACAAAAGAATGAACTGATTCAA  & 38268        \\
MLD24 & ATGAAAAAAATCAGCACAAACATACAATTCAATCTGCTGATTCAA  & 8429         \\
MLD25 & ATGAAAAAGATCACAACAAACAGGCAAAACAATGAACTGATTCAA  & 36120        \\
MLD27 & ATGAAAAAAATCGTCACAAACGAACAATTTAATGAACTGAAACAA  & 25949        \\
MLD28 & ATGAAAAAAATCGTGACAAACAGGCAATTTAATGAATTAAAACAA  & 22005        \\
MLD29 & ATGAAAAAAATCACACGAAAAGAACAATTTGAAGAACTGAAAAGG  & 36821        \\
MLD30 & ATGAAAAAAATCGTCAAGAACGAACAATTTAATGAACTAAAACAA  & 29816        \\
MLD31 & ATGAAAAAAATCACAACAAACGAACAATTTAATTATAAAAAACAA  & 10776        \\
MLD33 & ATGAAAAAAATCGTCACAAACGAACAATTTAATGAACTGAAACAA  & 26987        \\
MLD34 & ATGAAAAAAATCGTCACAAACGAACAAATAAATGAACTGAAACAA  & 14932        \\
MLD36 & ATGAAAAAAATCGTCGTAAACAGGCAAAACAATTATACAAAACAA  & 15280        \\
MLD37 & ATGAAAAAAATCGAGCAAAACAGGCAAAACAATTACTTAAAACAA  & 32678        \\
MLD38 & ATGAAAAAAATCGTCACAAACAGGCAAAACAATGAAAAAAAACAA  & 35159        \\
MLD39 & ATGAAAAAAATCACAACAAACGAACCTAACCAAAACCTGCCGCAA  & 10911        \\
MLD40 & ATGAAAAAAATCACAAATAACAGGCAAAACCAAACACTGAAACAA  & 52649        \\
MLD41 & ATGAAAAAAATCACACAGAACAGGAATAACCAAAATCTGAAACAA  & 16107        \\
MLD42 & ATGAAAAAAAAAACAACAAACAGGCAAAACCAAAATCTGAAACAA  & 49772        \\
MLD43 & ATGAAAAAAATCACAGTGAACAGGCAAAACCAAAATCTGAAACAA  & 15231        \\
MLD44 & ATGAAAAAAATCACACAGAACGGACAAAACCAAAACCTGAAAAGG  & 17563        \\
MLD45 & ATGAAAAAAATCACAACAAACAAAGTCAACCAAACTCTGAAACAA  & 32960        \\
MLD46 & ATGAAAAAAATCACAAACAACAGGCAAAACCAAACACTGAAACAA  & 45518        \\
MLD47 & ATGAAAAAAATCACAACAAACAAACAAAACAATGAAGTCAAACAG  & 51534        \\
MLD48 & ATGAAAAAAATCACAACAATTAGGCAAAACAATGAAACTAAAAGA  & 13777        \\
MLD49 & ATGAAAAAAAACACAAACAACAGGCAAAACCAGACACTGAAACAA  & 33663        \\
MLD50 & ATGAAAAAAAACACAAACAACAGGCAAAACCAGACACTAAAACAA  & 37441        \\
MLD51 & ATGAAAAAAAACACAAACAACAGGCAAAACCAAACACTCAAACAA  & 42801        \\
MLD52 & ATGAAAAAAATCACACGTGGCAGGCAAAACCAAACACTGAAACAA  & 18988        \\
MLD53 & ATGAAAAAAATTACAAACAAAAGGCAAAACCAAACACTGAAACAA  & 37895        \\
MLD54 & ATGAAAAAAAATTCGAACGAAAGGCAAAACCAAACACTGAAACAA  & 37426        \\
MLD55 & ATGAAAAAAAAGACAAACAACAGGCAAAACCAGACACTGAAACAA  & 36063        \\
MLD56 & ATGAAAAAAAACACAAACAACCGGCAAAACCAAACACTGAAACAA  & 38668        \\
MLD57 & ATGAAAAAAAATACAAACAACAGGCAAAACCAGACACTGAAACTC  & 32478        \\
MLD58 & ATGAAAAAAATAACAAACAAAAGGCAAAACCAAACACTGAAACAA  & 44898        \\
MLD59 & ATGAAAAAAATCACAAACAACATCCAAAACCAAACACTGAAACAA  & 35739        \\
MLD60 & ATGAAAAAAATCACAAACACAAGGCAAAACCAAACACTGAAATTG  & 26285        \\
MLD61 & ATGAAAAAAATCACAAACAATATCCAAAACCAAACACTGAAACAA  & 26565        \\
MLD62 & ATGAAAAAAATCACAAACAACAGGCAAAACCAAACACTGAAAGGT  & 70491        \\
MLD63 & ATGAAAAAAATCACAAACAACAGGCAAAACCAAACACTGAAATTG  & 33831        \\
MLD64 & ATGAAAAAAATCACAAACAACCGGCAAAACCAAACACTGAAACAA  & 44176        \\
MLD65 & ATGAAAAAAATCACAAACAACAGGCAAAACCAAACACTGAAACAA  & 45540        \\
MLD66 & ATGAAAAAAATCACAAACAACAGGCAAAACCAAACACTGAAACTT  & 33569        \\
MLD67 & ATGAAAAAAATCACAAACAACAGGCAAAACCAAACACTCAAACAA  & 43290        \\
MLD68 & ATGAAAAAAATCACAAACAACGGCCAAAACCAAACACTGAAACAA  & 23681        \\ \hline
\end{longtable}

\subsection{Further verification with 2-more experiment rounds}

Based on few-shot learning, our NCS design achieved its optimal solution within the first six rounds. We further added two additional rounds to see whether there's an improvement: 

\begin{figure}[ht]
    \centering
    \includegraphics[width = 1.0\linewidth]{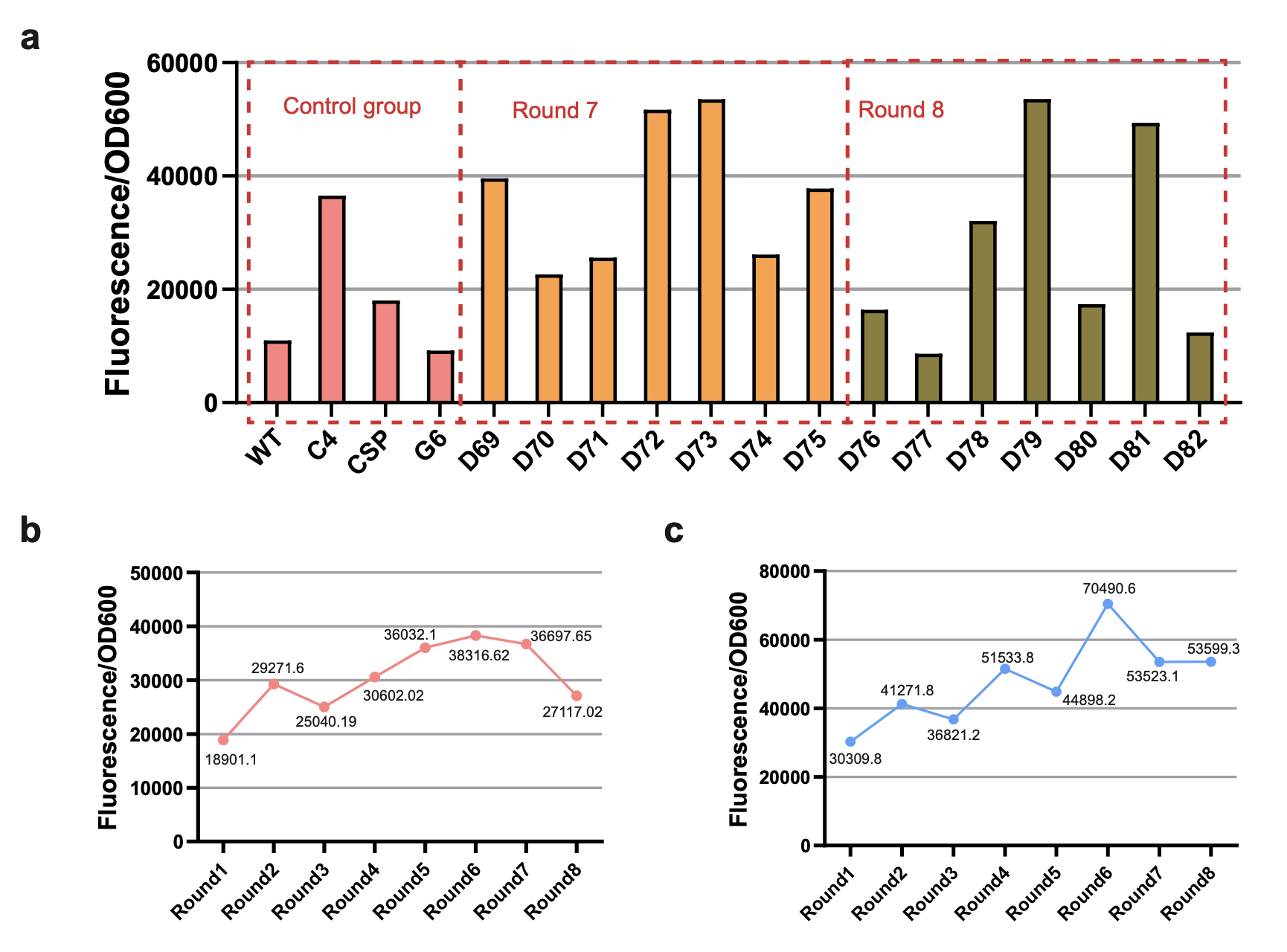}
        \caption{NCS expression intensity for two more addition rounds}

\begin{tablenotes}
\item {a) In the 7th and 8th iteration cycles, no stronger NCS emerged; b) The average strength of NCS per cycle initially showed a gradual increase, followed by a decline after the sixth round; c) The maximum strength of NCS in each cycle exhibited a fluctuating upward trend, reaching its peak in the sixth round }
\end{tablenotes}
    \label{ecoll}
\end{figure}

\subsection{Verification on Escherichia coli}
Building on our work with {\em B. subtilis}, we expanded our algorithm to explore its applicability in {\em Escherichia coli}, chosen for its well-characterized biology. As representatives of both Gram-positive and Gram-negative bacteria among prokaryotes, {\em B. subtilis} and {\em E. coli} exhibit distinct codon preferences. This divergence in codon usage implies that the control of gene expression by NCSs may differ between these two organisms. We tested six prominent genotypes—D40, D42, D46, D58, D62, and D65—as shown in Figure~\ref{ecoll}, in parallel experiments with {\em E. coli}.

Our results in Figure~\ref{ecoll} indicate that the NCS-designed genotypes raised expression levels by approximately 1.35-fold, differing notably from the $C_4$, csp, and G6 genotypes. Importantly, the experiment underlined that NCS strategies might not be universally transferable across species. The algorithm trained on {\em B. subtilis} showed constrained adaptability to {\em E. coli}. Future work will refine the model specifically for {\em E. coli} to optimize gene expression outcomes.

\begin{figure*}[ht]
    \centering
    \includegraphics[width = 0.5\linewidth]{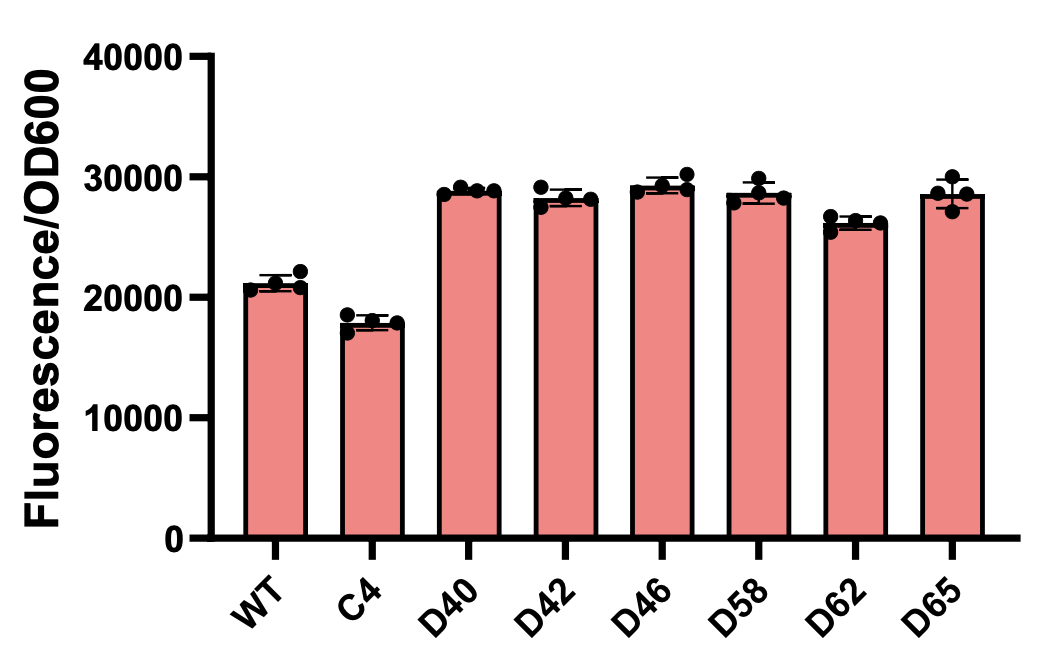}
        \caption{Parallel experiment for MLD-NCS in {\em E. coli}.}
    \label{ecoll}
\end{figure*}

\end{document}